\documentclass[prl,aps,twocolumn,preprintnumbers,showpacs]{revtex4}
\usepackage{latexsym,epsfig,amssymb,amsfonts,amsmath,graphicx,bbm,color,bm,times,hyperref,amstext,epstopdf}

\newcommand{\half}{\mbox{$\textstyle \frac{1}{2}$}}
\newcommand{\av}[1]{\langle #1\rangle}

\newcommand{\beq}{\begin{equation}}
\newcommand{\eeq}{\end{equation}}
\newcommand{\bea}{\begin{eqnarray}}
\newcommand{\eea}{\end{eqnarray}}

	\newcommand{\be}{\begin{equation}}
	\newcommand{\ee}{\end{equation}}

	\newcommand{\ket}[1]{{\left\vert {#1} \right\rangle}}	
	\newcommand{\bra}[1]{{\left\langle {#1} \right\vert}}

\allowdisplaybreaks

\begin{document}

\title{Extracting quantum work statistics and fluctuation theorems by single qubit interferometry}

\author{R. Dorner$^{1,2}$, S. R. Clark$^{2,3}$, L. Heaney$^{3}$, R. Fazio$^{3,4}$, J. Goold$^{2,5}$  and V. Vedral$^{2,3}$}
\address{$^1$Blackett Laboratory, Imperial College London, Prince Consort Road, London SW7 2AZ, United Kingdom}
\address{$^2$Clarendon Laboratory, University of Oxford, Parks Road, Oxford OX1 3PU, United Kingdom}
\address{$^3$Centre for Quantum Technologies, National University of Singapore, 3 Science Drive 2, Singapore 117543}
\address{$^4$NEST, Scuola Normale Superiore and Istituto Nanoscienze-CNR, I-56126 Pisa, Italy}
\address{$^5$Department of Physics, University College Cork, Cork, Ireland}

\date{\today}

\begin{abstract}
We propose an experimental scheme to verify the quantum non-equilibrium fluctuation relations using current technology. Specifically, we show that the characteristic function of the work distribution for a non-equilibrium quench of a general quantum system can be extracted from Ramsey interferometry of a single probe qubit. Our scheme paves the way for the full characterisation of non-equilibrium processes in a variety of quantum systems ranging from single particles to many-body atomic systems and spin chains. We demonstrate our idea using a time-dependent quench of the motional state of a trapped ion, where the internal pseudo-spin provides a convenient probe qubit.
\end{abstract}


\maketitle

{\it Introduction} -
In microscopic systems, non-equilibrium behaviour is dominated by thermal and quantum fluctuations. Obtaining a comprehensive understanding of these phenomena is therefore of great fundamental importance. In the past two decades, the discovery of the non-equilibrium fluctuation relations has made a significant contribution to this endeavour by characterising the full non-linear response of a microscopic system subject to a time-dependent force \cite{jrev,Jarzynski:97,jarzynski:04,Crooks}. Initially, these relations were derived for classical systems and experimentally confirmed in single-molecule stretching experiments \cite{colin}. More recently, their extension to the quantum regime has lead to theoretical progress in understanding the microscopic underpinnings of the laws of thermodynamics \cite{mrev,Tasaki}. Notably, however, an experimental verification of the quantum fluctuation relations is still forthcoming. 

In this letter, we propose an experimental scheme to extract the full statistics of work done in a non-equilibrium transformation of an arbitrary closed quantum system. 
The crux of our proposal is that the characteristic function of the work distribution can be extracted via Ramsey interferometry of a suitable probe qubit. Extracting the statistics of work in this way circumvents the requirement to perform projective energy measurements on the system of interest, as in previous proposals \cite{huber}, making our scheme generally applicable to a range of systems, including Bose \cite{Alessio:05,Bruderer:06} and Fermi gases \cite{goold:11, sindona:12}, spin chains \cite{rossini:07} and quenched ion strings \cite{Jens:11}. A further advantage of our scheme is that the temperature of the system is extracted directly from the measurement signal. This feature suggests potential future applications for in-situ, non-destructive thermometry in systems such as Bose-Einstein condensates.
We demonstrate the feasibility of our proposal using realistic parameters for the example of a trapped ion interacting with an external laser field. In this case, the internal pseudo-spin state of the ion provides a convenient probe qubit (see Ref.~\cite{mauro} for details of a similar scheme using hybrid opto-/electro-mechanical devices).

\noindent {\it Non-equilibrium quantum thermodynamics} -
We begin by briefly reviewing some important concepts from non-equilibrium
thermodynamics and defining the quantities that will be of interest in
the rest of this letter. We therefore consider a closed quantum
system described by a Hamiltonian $\hat
{H}(\lambda)$ containing an externally controlled parameter $\lambda(t)$. At time $t=0$ the control parameter has the initial value $\lambda(0)=\lambda_\textrm{i}$ and the system is prepared in the Gibbs state $\hat{\varrho}_\beta(\lambda_\textrm{i})=\textrm{exp}[-\beta \hat{H}(\lambda_\textrm{i})]/\mathcal{Z}_\beta(\lambda_\textrm{i})$, where $\mathcal{Z}_\beta(\lambda):=\textrm{tr}[\textrm{exp}(-\beta \hat{H}(\lambda))]$ is the partition function at inverse temperature $\beta$. The system is driven away from equilibrium by varying $\lambda(t)$ in a pre-defined, but otherwise arbitrary way over the quench time interval $t_\textrm{Q}$ to its final value $\lambda(t_\textrm{Q})=\lambda_\textrm{f}$. The initial and final Hamiltonians have the spectral decompositions $\hat{H}(\lambda_\textrm{i})=\sum_n \epsilon_n \ket{n}\bra{n}$ and $\hat{H}(\lambda_\textrm{f})=\sum_m \bar{\epsilon}_m \ket{\bar{m}}\bra{\bar{m}}$, respectively, and the protocol $\hat{H}(\lambda_\textrm{i}) \rightarrow \hat{H}(\lambda_\textrm{f})$ that connects them generates the unitary evolution $\hat{U}(t_\textrm{Q})$.

The work done on the system $W$ is defined by two projective energy measurements: The first, at $t=0$, projects onto the eigenbasis of $\hat{H}(\lambda_\textrm{i})$ and gives the outcome $\epsilon_n$ with a probability $p_n=\exp(-\beta \epsilon_n)/\mathcal{Z}_\beta(\lambda_\textrm{i})$. The second measurement, at the end of the quench protocol $t=t_\textrm{Q}$, projects onto the eigenbasis of $\hat{H}(\lambda_\textrm{f})$ and gives the outcome $\bar{\epsilon}_m$ with probability $p_{m|n}=|\bra{\bar{m}}\hat{U}(t_\textrm{Q})\ket{n}|^2$. The \textit{quantum work distribution}, which encodes the random fluctuations in non-equilibrium work arising from both thermal ($p_n$) and quantum measurement ($p_{m|n}$) statistics over many identical realisations of the quench protocol, is thus,
\begin{align}
P_\textrm{F}(W):=\sum_{n,m}p_n p_{m|n}\delta(W-(\bar{\epsilon}_m-\epsilon_n)).
\nonumber
\end{align}
Here `$\textrm{F}$' denotes that this is the work distribution for the \textit{forward} process $\hat{H}(\lambda_\textrm{i}) \rightarrow \hat{H}(\lambda_\textrm{f})$. The corresponding \textit{backward} work distribution $P_\textrm{B}(W)$ is obtained by preparing the system in the Gibbs state $\hat{\varrho}_\beta(\lambda_\textrm{f})$ of the final Hamiltonian and subjecting it to the time-reversed protocol $\hat{H}(\lambda_\textrm{f}) \rightarrow \hat{H}(\lambda_\textrm{i})$ generated by the evolution $\hat{U}^\dagger(t_\textrm{Q})$.

By studying the fluctuations in non-equilibrium work it is possible to extract important {\it equilibrium} information. This is revealed by the non-equilibrium fluctuation relations, in particular, the Tasaki-Crooks relation \cite{Crooks,Tasaki}
\begin{align}
\frac{P_\textrm{F}(W)}{P_\textrm{B}(-W)}=e^{\beta(W-\Delta F)},
\label{crooks}
\end{align}
which shows that, for any closed quantum system undergoing an arbitrary non-equilibrium transformation, the fluctuations in work are related to the equilibrium free energy difference $\Delta F=(1/\beta)\textrm{ln}[\mathcal{Z}_\beta(\lambda_\textrm{i})/\mathcal{Z}_\beta(\lambda_\textrm{f})]$. This relationship is further emphasized by a corollary to Eq.~\eqref{crooks} known as the Jarzynski equality \cite{Jarzynski:97};
\begin{align}
\int \textrm{d}W P_\textrm{F}(W)e^{-\beta W}=\langle e^{-\beta W} \rangle=e^{-\beta \Delta F},
\nonumber
\end{align} 
which states that $\Delta F$ can also be extracted from the properties of the forward (or backward) work distribution alone.

The primary quantities of interest in this letter are the forward and backward \emph{characteristic functions}, defined as the Fourier transform of their corresponding work distribution \cite{lutz}. Hence, the forward characteristic function  (taking $\hbar = 1$)
\begin{align}
\chi_\textrm{F}(u)&:=\int \textrm{d}W e^{iuW}P_\textrm{F}(W),
\nonumber \\
&=\textrm{tr}[\hat{U}^\dagger(t_\textrm{Q})e^{iu\hat{H}(\lambda_\textrm{f})}\hat{U}(t_\textrm{Q})e^{-iu\hat{H}(\lambda_\textrm{i})}\hat{\varrho}_\beta(\lambda_\textrm{i})],
\label{charfunction}
\end{align}
while the backward characteristic function $\chi_\textrm{B}(u):=\int \textrm{d}W e^{iuW}P_\textrm{B}(W)$.

Previous experimental proposals to extract the full statistics of work, and hence verify the quantum fluctuation theorems, have sought to directly measure the work distribution via a series of projective energy measurements \cite{huber}. However, even for quantum systems of modest complexity this can be practically challenging. In the following section, we show how this difficulty can be avoided by instead extracting the characteristic function of the work distribution using well-established experimental techniques.

{\it Experimental extraction of the characteristic function} - 
The purpose of our proposal is to measure the work done in a non-equilibrium transformation of a generic quantum system by temporarily coupling it to an easily-addressable probe qubit. We assume that the total Hamiltonian describing the probe qubit and system of interest has the form $\hat{H}_\textrm{T}(t)=\frac{\Delta}{2}\hat{\sigma}_z + \hat{H}_\textrm{S} + \hat{H}_\textrm{I}(t)$ where $\Delta$ is the splitting between the ground $\ket{\downarrow}$ and excited $\ket{\uparrow}$ states of the qubit, which are eigenstates of the spin-1/2 Pauli-$z$ operator $\hat{\sigma}_z$ (similarly $\hat{\sigma}_x$ and $\hat{\sigma}_y$ denote the Pauli-$x$ and -$y$ operators) and $\hat{H}_\textrm{S}$ is the time-independent bare Hamiltonian of the system of interest. The qubit-system interaction term $\hat{H}_\textrm{I}(t)$ contains all of the time-dependence and is assumed to have the form
\begin{align}
\hat{H}_\textrm{I}(t)=\Big(g_\downarrow(t) \ket{\downarrow}\bra{\downarrow}+ g_\uparrow(t)\ket{\uparrow}\bra{\uparrow}\Big)\otimes\hat{V},
\nonumber
\end{align}
where $g_\uparrow(t)$ and $g_\downarrow(t)$ are externally controlled parameters and $\hat{V}$ is a perturbation acting on the system of interest. With an interaction of this form, an isolated quench of the system described by the protocol $\hat{H}(\lambda_\textrm{i})=\hat{H}_\textrm{S}+\lambda_\textrm{i}\hat{V} \rightarrow \hat{H}(\lambda_\textrm{f})=\hat{H}_\textrm{S}+\lambda_\textrm{f}\hat{V}$  can be realised by varying both of the spin-dependent parameters, $g_\downarrow(t)$ and $g_\uparrow(t)$, from the initial values $g_\downarrow(0)=g_\uparrow(0)=\lambda_\textrm{i}$ at $t=0$ to the final values $g_\downarrow(t_\textrm{Q})=g_\uparrow(t_\textrm{Q})=\lambda_\textrm{f}$ at $t=t_\textrm{Q}$ according to the same, pre-defined protocol.

To extract the characteristic function for this quench, however, a related but distinct extraction protocol must be performed on the system \textit{and} qubit as part of a modified Ramsey sequence in which $g_\downarrow(t)$ and $g_\uparrow(t)$ are varied over the Ramsey time interval $t_\textrm{R}$. Crucially, during the extraction protocol, the parameters $g_\downarrow(t)$ and $g_\uparrow(t)$ will differ so that the probe qubit and system become coupled. Explicitly, the experimental procedure to extract the characteristic function is as follows: \textit{i)} For times $t\leq0$  the qubit is decoupled from the system by holding the spin-dependent couplings fixed at $g_\downarrow(0)=g_\uparrow(0)=\lambda_\textrm{i}$. Furthermore, the qubit and system are thermalised in the product state $\hat{\rho}=\ket{\downarrow}\bra{\downarrow}\otimes \varrho_\beta(\lambda_i)$ by ensuring that $\beta\Delta \gg 1$. \textit{ii}) At $t=0$, a Hadamard operation $\hat{\sigma}_\textrm{H}=(\hat{\sigma}_x+\hat{\sigma}_z)/\sqrt{2}$ is applied to the qubit. \textit{iii}) The spin-dependent couplings are \textit{independently} varied over the Ramsey time internal $t_\textrm{R}$ from their initial value $g_\downarrow(0)=g_\uparrow(0)=\lambda_\textrm{i}$ to the final value $g_\downarrow(t_\textrm{R})=g_\uparrow(t_\textrm{R})=\lambda_\textrm{f}$. This protocol generates the unitary evolution operator $\hat{T}(t_\textrm{R})$ that acts in the joint Hilbert space of the qubit and system to generate a conditional dynamical quench of the system contingent upon the state of the probe qubit. For the sake of clarity, we stress that this protocol is \textit{not} the same protocol for which we wish the extract the work done. \textit{iv}) At the end of the protocol, $t=t_\textrm{R}$, the qubit and system are automatically decoupled and a second Hadamard operation is applied to the qubit at.

\begin{figure}[t!]
\begin{center}
\includegraphics{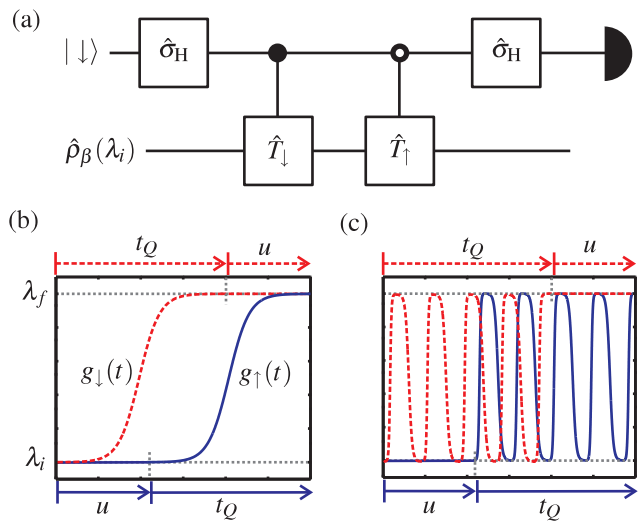}
\caption{(a) The Ramsey sequence represented as a quantum circuit. The probe qubit in the upper branch is prepared in the $\ket{\downarrow}$ state and the system of interest is prepared in the state $\varrho_\beta(\lambda_i)$ defined in the text. A black (white) circle indicates that the operation is controlled on the probe qubit being in the $\ket{\downarrow}$ ($\ket{\uparrow}$) state. In (b) and (c) examples are shown of the time variation of the spin couplings $g_\downarrow(t)$ and $g_\uparrow(t)$ over the Ramsey scheme time $t_R = t_Q + u$ required to obtain the characteristic function $\chi_\textrm{F}(u)$. (b) A forward process of the form $\lambda(t)=\lambda_i + (\lambda_f - \lambda_i)[1 + \textrm{tanh}(t/T)]/2$, where $T$ is the switching time and the total quench time is $t_\textrm{Q}=8T$ is shown. (c) A forward process is shown which is composed of repeated fast and slow tanh switching between $\lambda_i$ and $\lambda_f$, like that in (b), before ending at $\lambda_f$.}
\label{fig1}
\end{center}
\end{figure}

The output state of the probe qubit at the end of the Ramsey sequence is, thus,
\begin{align}
\hat{\rho}_\textrm{q} = {}&\textrm{tr}_\textrm{S}\left[\hat{\sigma}_\textrm{H} \hat{T}(t_\textrm{R}) \hat{\sigma}_\textrm{H} \hat{\rho} \,\hat{\sigma}_\textrm{H} \hat{T}^\dagger(t_\textrm{R})\hat{\sigma}_\textrm{H} \right]
\nonumber \\
={}&\frac{1+\Re\left[L(t_\textrm{R})\right]}{2}\ket{\downarrow}\bra{\downarrow}+\frac{i\Im\left[L(t_\textrm{R})\right]}{2}\ket{\downarrow}\bra{\uparrow}
\nonumber \\
&- \frac{i\Im\left[L(t_\textrm{R})\right]}{2}\ket{\uparrow}\bra{\downarrow}
+\frac{1-\Re\left[L(t_\textrm{R})\right]}{2}\ket{\uparrow}\bra{\uparrow},
\label{ramsey}
\end{align}
where we have introduced the \textit{decoherence factor}
\begin{align}
L(t_\textrm{R})=\textrm{tr}_\textrm{S}
[\hat{T}^\dagger_\uparrow(t_\textrm{R})\hat{T}_\downarrow(t_\textrm{R}) \hat{\varrho}_\beta(\lambda_\textrm{i})].
\label{lecho}
\end{align}
Here, the unitary operators $\hat{T}_\downarrow(t_\textrm{R})=\bra{\downarrow}\hat{T}(t_\textrm{R})\ket{\downarrow}$ and $\hat{T}_\uparrow(t_\textrm{R})=\bra{\uparrow}\hat{T}(t_\textrm{R})\ket{\uparrow}$ act in the Hilbert space of the system of interest and describe its evolution under the two different time-dependent quenches generated by $g_\downarrow(t)$ and $g_\uparrow(t)$, respectively. Consequently, the Ramsey sequence (shown in Fig.~\ref{fig1}(a)) creates an entangled state between the basis states of the probe qubit and the two quenched states of the system, $\hat{T}_\downarrow [\varrho_\beta(\lambda_\textrm{i})] \hat{T}^\dagger_\downarrow$ and $\hat{T}_\uparrow [\varrho_\beta(\lambda_\textrm{i})] \hat{T}^\dagger_\uparrow$. The real $\Re [L(t_\textrm{R})]$ and imaginary $\Im[L(t_\textrm{R})]$ parts of the decoherence factor define the populations and coherences of the probe qubit density matrix in Eq.~\eqref{ramsey} and are experimentally reconstructed by measuring $\hat{\sigma}_z$ and $\hat{\sigma}_y$ over many identical experimental runs. 

A direct relationship between $L(t_\textrm{R})$ in the Ramsey scheme and the characteristic function $\chi_\textrm{F}(u)$ for the quench protocol $\hat{H}(\lambda_\textrm{i})=\hat{H}_\textrm{S}+\lambda_\textrm{i}\hat{V} \rightarrow \hat{H}(\lambda_\textrm{f})=\hat{H}_\textrm{S}+\lambda_\textrm{f}\hat{V}$  is established by judiciously engineering the unitary operators $\hat{T}_\downarrow(t_\textrm{R})$ and $\hat{T}_\uparrow(t_\textrm{R})$; As illustrated in Fig.~\ref{fig1}(b)-(c), the spin-dependent control parameters $g_\downarrow(t)$ and $g_\uparrow(t)$ are varied so that the system state entangled with $\ket{\uparrow}$ undergoes the quench $\lambda_\textrm{i} \rightarrow \lambda_\textrm{f}$ over the time interval $t_\textrm{Q}$, followed by constant evolution at $\lambda_\textrm{f}$ up to the Ramsey time $t_\textrm{R}$. Simultaneously, the system state entangled with $\ket{\uparrow}$ undergoes constant evolution at $\lambda_\textrm{i}$ followed by the quench. The corresponding unitary operators are thus
\begin{align}
\hat{T}_\uparrow(t_\textrm{R})&=e^{-i(t_\textrm{R}-t_\textrm{Q})\hat{H}(\lambda_\textrm{f})}\hat{U}(t_\textrm{Q}),
\nonumber \\
\hat{T}_\downarrow(t_\textrm{R})&=\hat{U}(t_\textrm{Q})e^{-i(t_\textrm{R}-t_\textrm{Q})\hat{H}(\lambda_\textrm{i})},
\nonumber
\end{align}
which, after identifying $u=t_\textrm{R}-t_\textrm{Q}$, show that the decoherence factor in Eq.~\eqref{lecho} coincides \textit{exactly} with the forward characteristic function in Eq.~\eqref{charfunction}. Hence, the characteristic function is extracted by embedding the evolution $\hat{U}(t_\textrm{Q})$ into the qubit-system evolution and repeating the protocol for different  run times $t_\textrm{R} \geq t_\textrm{Q}$. The corresponding backwards characteristic function is obtained by a straightforward modification of the above scheme. In both cases, the work distributions $P_\textrm{F}(W)$ and $P_\textrm{B}(W)$ are obtained from the inverse Fourier transform of their respective characteristic functions (cf. Eq.~\eqref{charfunction}). In the following section we illustrate how this scheme can be implemented using a conventional trapped ion arrangement under realistic circumstances.

{\it Implementation using a trapped ion} - We consider a single ion of mass $M$ contained in a linear Paul trap~\cite{ionreview}. By using the $S_{1/2}$ ground state Zeeman sublevels of the ion $\ket{m=1/2} = \ket{\uparrow}$ and $\ket{m=-1/2}=\ket{\downarrow}$ this system provides an ideal realization of a spin-$1/2$ particle confined in a harmonic potential. We therefore have $\hat{H}_\textrm{S}= \omega_0 (\hat{a}^\dagger \hat{a}+1/2)$, where $\omega_0$ is the natural frequency of the oscillator and $\hat{a}^\dagger$ ($\hat{a}$) is the oscillator raising (lowering) operator. The trapped ion setup has a number of distinguishing features: First, accurate detection of the ion's internal states can be accomplished by observing the scattered fluorescence from near-resonant driving of a cycling transition. Second, transformations between internal states can be implemented by a Raman transition (e.g. performing the Hadamard operation $\sigma_\textrm{H}$ via a $\pi/2$ pulse) and the tunable azimuthal phase of the transition permits both $\av{\hat{\sigma}_z}$ and $\av{\hat{\sigma}_y}$ to be determined from the fixed final measurement~\cite{ionreview}. Third, precise preparation of the initial thermal state $\hat{\varrho}_\beta$, with mean phonon number $\bar{n} = [\exp(\beta\omega_0) - 1]^{-1}$, can be achieved by allowing heating after resolved-sideband laser cooling to the motional ground state or Doppler cooling on the $S_{1/2}$ to $P_{1/2}$ transition~\cite{ionreview,huber}. 

Similar to an earlier proposal studying quantum chaos using a trapped ions~\cite{zollerchaos}, we quench the motional state of the ion by illuminating it with a far-detuned elliptically polarized standing wave laser field. Further, since the $\sigma^+$ and $\sigma^-$ polarized contributions couple exclusively to the $\ket{\downarrow}$ and $\ket{\uparrow}$ states, respectively, they induce a spin-dependent optical dipole potential for the ion~\cite{mcdonnell}. Hence, after making the rotating-wave approximation and adiabatically eliminating the far-detuned excited states, we find the interaction Hamiltonian~\footnote{We have absorbed the Stark shifts for the $\ket{\uparrow}$ and $\ket{\downarrow}$ states into the qubit splitting $\Delta$.}
\begin{align}
\hat{H}_\textrm{I}=\Big(g_\downarrow(t) \ket{\downarrow}\bra{\downarrow}+g_\uparrow(t)\ket{\uparrow}\bra{\uparrow}\Big)\otimes\sin^2(k\hat{x} + \phi),
\nonumber
\end{align}
where $k$ is the magnitude of the wave-vector orientated along the axis of the trap for both polarisations, and $\phi$ is the phase of the standing waves relative to the trap centre at $x=0$. The couplings $\Omega_{\uparrow}(t)$ and $\Omega_{\downarrow}(t)$ are the time-dependent Rabi frequencies, which are independently controlled by varying the laser intensity for the corresponding polarization. 

In the Lamb-Dicke regime, quantified by $\eta = k x_0 \ll 1$ where $x_0 = (2M\omega_0)^{-1/2}$, the extent of the ions motion is small compared to the spatial variation of the optical dipole potential. Consequently, expanding $\hat{H}_\textrm{I}(t)$ to $O(\eta^3)$ around $x=0$ gives an energy shift $\epsilon_\sigma(t) = \Omega_\sigma(t) \sin^2(\phi)$, a linear potential of strength $g_\sigma(t) = \eta\Omega_\sigma(t)\sin(2\phi)$ and frequency change $\tilde{\omega}_\sigma= \omega_0 + 4 \eta^2 \Omega_\sigma(t) \cos(2 \phi)$, where $\sigma = \{\uparrow,\downarrow\}$. Accordingly, by choosing the appropriate relative phase, the optical dipole potential can cause the oscillator to be tightened ($\phi=0$), slackened ($\phi=\pi/2$) or displaced ($\phi=\pi/4$), while other phases lead to combinations of these effects. For concreteness, we focus on a pure displacement quench where the perturbation reduces to $\hat{V} = x_0(\hat{a}^\dagger + \hat{a})$ in addition to the energy shift $\epsilon_\sigma(t)$. Since $g_\sigma(t) \propto \Omega_\sigma(t)$ the protocol can be implemented by varying the laser intensities of the two orthogonally polarized standing waves. 

\begin{figure}[t!]
\begin{center}
\includegraphics{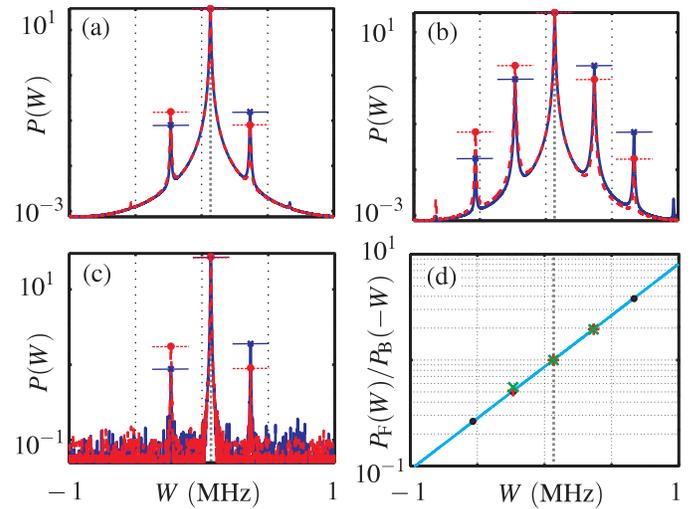} 
\caption{Here a $^{40}$Ca$^+$ is assumed to be confined with an axial trapping frequency $\omega_0 = 300$ kHz and in an initial thermal state with $\bar{n} =1$. The standing wave optical dipole potential for either polarization are taken as being generated by a 397 nm laser $\approx 30$ GHz detuned from the $S_{1/2}-P_{1/2}$ transition, giving $\eta = 0.33$, and with a maximum Rabi frequency of $\Omega = 150$ kHZ~\cite{mcdonnell,poschinger}. A sampling rate of 2 MHz, along with a decoherence time scale of $\tau = 15\times (2\pi/\omega_0) = 50$ $\mu$s has been used and measurements were performed up to a time $\approx 500$ $\mu$s where signal was completely damped. (a) The quantum work distributions $P_\textrm{F}(W)$ (solid) and $P_\textrm{B}(-W)$ (dashed) are plotted on a logarithmic scale for a single $\textrm{tanh}$ switching of $\lambda(t)$, as shown in Fig.~\ref{fig1}(b), with $T = 0.3 \times (2\pi/\omega_0) = 1$ $\mu$s. The distributions are continuous due to decoherence induced broadening. The dots with horizontal lines denote the peak amplitudes identified from both $P_\textrm{F}(W)$ and $P_\textrm{B}(-W)$. The vertical dashed line is at a frequency of 67 kHz corresponding to the exact $\Delta F$. (b) The plot as (a) but with a repeated $\textrm{tanh}$ switching quench (as shown in Fig.~\ref{fig1}(c)) with $T_{\rm{fast}} = 0.2 \times (2\pi/\omega_0) = 0.03$ $\mu$s and $T_{\rm{slow}} = 3 \times (2\pi/\omega_0) = 20$ $\mu$s. (c) The plot and setup as (b) with $0.5\%$ Gaussian noise added to the signals $\chi_\textrm{F}(u)$ and $\chi_\textrm{B}(u)$. (d) The ratio of the $P_\textrm{F}(W)/P_\textrm{B}(-W)$ against $W$ evaluated from the peaks identified in (a) ($+$), (b) ($\bullet$) and (c) ($\times$) plotted on a logarithmic scale. The solid line is Crooks relation from Eq.~\eqref{crooks} expected from the exact $\beta$ and $\Delta F$. A best fit of the function $\exp(AW - B)$ for the noisy spectrum in (c) gives $A = 0.72/\omega_0$ and $B/A = 0.20\omega_0$, which compare to the exact values $\beta = 0.69/\omega_0$ and $\Delta F = 0.22\omega_0$, respectively.}
\label{fig2}
\end{center}
\end{figure}

To examine the verification of the Crooks relation in Eq.~\eqref{crooks} we have numerically computed $\chi_\textrm{F}(u)$ and $\chi_\textrm{B}(u)$ for two different quenches in a possible $^{40}$Ca$^+$ ion experiment~\cite{mcdonnell,poschinger}. Two experimental limitations were modelled in this calculation; First, we used a realistic sampling rate for the measurement of $\chi_\textrm{F}(u)$ to account for discrete data. Second, an enveloping factor $\exp(-u/\tau)$, with a decay time $\tau$, was added to the measurement signal to account for decoherence of the entangled state that appears within the scheme~\cite{turchette}. In Fig.~\ref{fig2}(a) the resulting quantum work distributions $P_\textrm{F}(W)$ and $P_\textrm{B}(W)$ are shown for a simple $\textrm{tanh}$ forward quench from $\lambda_\textrm{i} = 0$, like that shown in Fig.~\ref{fig1}(b), described by $\lambda(t) = \lambda_\textrm{f}[1 + \textrm{tanh}(t/T)]/2$. The switching time-scale $T$ used was a fraction of the natural trap frequency $2\pi/\omega_0$. Both $P_\textrm{F}(W)$ and $P_\textrm{B}(W)$ are composed of $\delta$-peaks, separated by $\omega_0$ and broadened into a continuous spectrum by the decoherence envelope. Once inverted, as $P_\textrm{B}(-W)$, the peaks in the two spectra line up. As the quench is non-adiabatic, the first-order peaks are visible, though much weaker than the carrier peak. In Fig.~\ref{fig2}(b) the same spectra are plotted for a quench composed of repeated fast and slow $\textrm{tanh}$ switches (see Fig.~\ref{fig1}(c)) with the fast switching on the order of a hundredth of $2\pi/\omega_0$. The stronger first-order peaks and now-visible second-order peaks evidence how this quench is more non-adiabatic. In Fig.~\ref{fig2}(c) we model an additional limitation by computing the spectra for the same quench as Fig.~\ref{fig2}(b) with $0.5\%$ Gaussian noise added to $\chi_\textrm{F}(u)$ and $\chi_\textrm{B}(u)$. Despite this white-noise the first-order peaks remain visible.
 
The analysis is completed by extracting the amplitudes from $P_\textrm{F}(W)$ and $P_\textrm{B}(W)$ of all identifiable peaks. The ratio $P_\textrm{F}(W)/P_\textrm{B}(-W)$ is then computed for these energies $W$. The Crooks relation in Eq.~\eqref{crooks} predicts that they will lie on a exponential curve, which for the exact $\beta$ and $\Delta F$ of the quench is shown on a logarithmic scale in Fig.~\ref{fig2}(d) as the solid line. The extracted ratios for the three examples given in Fig.~\ref{fig2}(a)-(c) are also plotted and found to cluster tightly on this line. For each case, a fitting to $\exp(A W - B)$ is made with the parameter $A$ providing an estimate of $\beta$, thereby establishing that the interferometric protocol also acts as a thermometer. This value can be independently compared to a direct measurement of the initial phonon distribution of the ion. Using $A$, the fit parameter $B/A$ subsequently allows an estimate of $\Delta F$ to be extracted~\cite{free}. For the cases in Fig.~\ref{fig2}(a)-(b) the fits essentially yield the exact result $\Delta F = \epsilon_\sigma(t_R) - \half g_\sigma(t_R)^2/\omega_0$, demonstrating the quench independence of the relation Eq.~\eqref{crooks}. The noisy spectra in Fig.~\ref{fig2}(c) also provides a good estimate (see caption) of both $\beta$ and $\Delta F$ from its zeroth- and first-order peaks, illustrating a degree of robustness in the scheme when sufficiently non-adiabatic quenches are employed. 

{\it Discussions and conclusions} -
We have outlined a general experimental scheme to extract the full statistics of work done on a quantum system.
Our scheme uses Ramsey interferometry of a single probe qubit to extract the characteristic function of the work distribution following an non-equilibrium quench of a quantum system. This bypasses the requirement of other proposals to implement resource intensive projective energy measurements, with the added benefit of being generally applicable to a wide range of current quantum technologies. We have demonstrated the feasibility of our scheme using a conventional ion-trap system and standard tools for laser manipulation under realistic conditions. As such our proposal should pave the way for the first experimental verification of fluctuation relations in the quantum regime. Beyond this, we propose that our work is easily adjusted to probe manybody systems where recent studies have shown that the statistics of work can be used to shed light on the universal critical features of models from many-body physics \cite{Silva, dorner, Silva2}. In addition further studies have also established an intriguing relationship between energy fluctuations in a local quench and the block entanglement of a many-body state \cite{Cardy} and it has been suggested that these ideas can be tested by a local probe \cite{demler_switch}. We therefore suggest that the scheme presented here may be used to probe non-local correlations in a macroscopic out-of equilibrium many-body system.

{\it Acknowledgements} - The authors thank M. Paternostro for informative discussions relating to the topic of this manuscript. They also thank U. Poschinger, A. Alberti and S. Deffner for helpful comments on an earlier version of this manuscript. RD is funded by the EPSRC; JG acknowledges
funding from IRCSET through a Marie Curie
International Mobility fellowship; SRC, LH, RF and VV thank the National Research Foundation and the
Ministry of Education of Singapore for support; VV is also is a fellow of Wolfson College Oxford
and is supported by the John Templeton Foundation and the Leverhulme Trust (UK).


\begin{thebibliography}{100}
\bibitem{free} The analytic solution for a displaced harmonic oscillator reveals that the carrier peak of the work distributions is always located precisely at $\Delta F$, and will consequently have an identical amplitude in both spectra. More general perturbations will not possess this property and are instead analysed by the method described.
\end{thebibliography}
\end{document}